\documentclass[%
reprint,
 amsmath,amssymb,
 aps,
]{revtex4-1}

\usepackage{graphicx}
\usepackage{dcolumn}
\usepackage{bm}

\begin{document}

\title{Supersymmetric quantum mechanics requires $g=2$ for vector bosons}

\author{Georg Junker}
 \altaffiliation[Also at ]{Institut f\"ur Theoretische Physik I, Universit\"at Erlangen-N\"urnberg, Staudtstra{\ss}e 7, D-91058 Erlangen, Germany}
 \email{\\ gjunker@eso.org}
\affiliation{%
 European Southern Observatory, Karl-Schwarzschild-Stra{\ss}e 2, D-85748 Garching, Germany
}
\homepage{http://www.eso.org/~gjunker}
\date{\today}

\begin{abstract}
Relativistic arbitrary spin Hamiltonians are shown to obey the algebraic structure of supersymmetric quantum system if their odd and even parts commute. This condition is identical to that required for the exactness of the Foldy-Wouthuysen transformation. Applied to a massive charged spin-$1$ particle in a constant magnetic field, supersymmetric quantum mechanics  necessarily requires a gyromagnetic factor $g=2$.
\end{abstract}

\keywords{Relativistic wave equations, Supersymmetry, g factor}

\maketitle

\section{Introduction}
There is some common agreement that the gyromagnetic ratio $g$ of charged elementary particles when coupled to an electromagnetic field is $g=2$.  There are several reasonable arguments for this. The equation of motion for the spin vector, as shown by Bargmann, Michel and Telegdi \cite{BMT1959}, takes a very simple form when $g=2$. As argued by Weinberg \cite{Wein1970} and much later by Ferrara, Porrati  and Telegdi \cite{FPT1992}, the requirement to have a good high-energy behaviour of scattering amplitudes, one must choose $g=2$ for any spin. In 1994 Jackiw \cite{Jackiw1998} showed that the value $g=2$ also follows from a gauge symmetry in the case of a massless spin-$1$ field coupled to an electromagnetic field.
This is of course in agreement with the standard model. Here the currently known electrically charged elementary particles are either spin-$\frac{1}{2}$ fermions or spin-$1$ bosons. The standard model indeed requires for all these charged particles a value $g=2$ at the tree level. Whereas for the elementary fermions ($s=1/2$) this assertion is consistent with the Belifante \cite{Belifante1953} conjecture $g=1/s$, it obviously disagrees with the case of vector bosons where $s=1$ and the Belifante conjecture would imply $g=1$. In fact, precision measurements at the Tevatron \cite{Kotwal} resulted in the bounds $1.944 \leq g \leq 2.080$ at $95\%$ C.L.\ for the W boson.
Hence, the case of spin-$1$ elementary particles is of particular interest as no massive and charged higher-spin elementary particles are known yet.

In 1992 Ferrara and Porrati \cite{FP1992} utilized an $N=1$ supersymmetry algebra to show that "the gyromagnetic ratio of arbitrary-spin supersymmetric particles must be equal to 2". Whereas Ferrara and Porrati considered a supersymmetry (SUSY) where the supercharges transform bosons into fermions and vice versa, we consider in this brief report relativistic Hamiltonians for arbitrary spin, that is, characterising the relativistic dynamics of a standard particle with arbitrary but fixed spin. Under the assumption that the even and odd part of such a  Hamiltonian commute, it is shown that one can construct a SUSY structure know from SUSY quantum mechanics. Here the SUSY charge transforms between positive and negative energy eigenstates. For this we first generalise the concept of a supersymmetric Dirac Hamiltonian to relativistic Hamiltonians for arbitrary spin $s=0,\frac{1}{2},1,\frac{3}{2},\ldots$. Then we consider the case of a charged particle with $s=1$ interacting with an external constant magnetic field. Here the coupling to the spin-degree of freedom is a prior considered with arbitrary gyromagnetic ration $g$. It is shown that supersymmetric quantum mechanics requires $g=2$.

\section{Supersymmetric Relativistic Hamiltonians}
The Hamiltonian of an arbitrary spin-$s$ Hamiltonian can be put into the form \cite{Foldy1956}
\begin{equation}\label{1}
  H=\beta {\cal M} + {\cal O}\,,
\end{equation}
which acts on the Hilbert space ${\cal H}=L^2(\mathbb{R}^3)\otimes\mathbb{C}^{2(2s+1)}$ whose elements are $2(2s+1)$-dimensional spinors. The matrix $\beta$ obeys the relation $\beta^2 = 1$ and may be represented as a block-diagonal matrix
\begin{equation}\label{2}
  \beta = \left( \begin{array}{cc} 1 & 0 \\ 0 & -1 \end{array} \right)
\end{equation}
where here the $1$ stands for the $2s+1$-dimensional unit matrix. Let us note that the two parts of the Hamiltonian (1) are chosen such that ${\cal M}$ accommodates all the even elements and  ${\cal O}$ all the odd elements of $H$ with respect to $\beta$. That is, we have the commutation and anti-commutation relations
\begin{equation}\label{3}
  [\beta, {\cal M}]=0\,, \quad\{\beta, {\cal O}\}=0\,.
\end{equation}
Note that the Hamiltonian (1) is hermitian, i.e. $H=H^\dag$, only for the case of fermions, and hence for half-odd-integer $s$. For bosons, where $s$ is integer, the Hamiltonian is pseudo-hermitian, i.e. $H=\beta H^\dag \beta$. Having this in mind it is straightforward to show that in the representation (2) both parts of $H$ are necessarily of the form
\begin{equation}\label{4}
  {\cal M} = \left(\begin{array}{cc}
                     M_+ & 0 \\ 0 & M_- \end{array}\right)\,,\quad
  {\cal O} = \left(\begin{array}{cc}
                     0 & A \\ (-1)^{2s+1}A^\dag & 0 \end{array}\right)\,,
\end{equation}
where $M_\pm:{\cal H}_\pm\mapsto{\cal H}_\pm$ with $M^\dag_\pm= M_\pm$, $A:{\cal H}_-\mapsto{\cal H}_+$ and $A^\dag:{\cal H}_+\mapsto{\cal H}_-$. Here we have introduced the positive and negative energy subspaces ${\cal H}_+$ and ${\cal H}_-$, which are also eigenspaces of $\beta$ for eigenvalue $+1$ and $-1$, respectively. Note that ${\cal H}={\cal H}_+\oplus{\cal H}_-$ and ${\cal H}_\pm=L^2(\mathbb{R}^3)\otimes\mathbb{C}^{2s+1}$. Well-know examples are the Klein-Gordon $(s=0)$ and the Dirac ($s=1/2$) particle in a magnetic field \cite{Feshbach1958}.

Let us now assume that the odd and even parts of $H$ commute, that is, $[{\cal M}, {\cal O}]=0$. This assumption implies
\begin{equation}\label{5}
  M_+ A = A M_-\,,\qquad  A^\dag M_+ = M_- A^\dag\,.
\end{equation}
From this condition follows that the squared Hamiltonian (1) becomes block diagonal as the off-diagonal blocks vanish due to (5). That is
\begin{equation}\label{6}
  H^2 = \left(\begin{array}{cc} M^2_+ + (-1)^{2s+1} AA^\dag& 0 \\ 0 & M^2_- +(-1)^{2s+1} A^\dag A \end{array}\right)\,,
\end{equation}
which allows us to define a SUSY structure in analogy to the Dirac case \cite{Thaller1992,Junker2019}. To be more explicit let us introduce the non-negative SUSY Hamiltonian
\begin{equation}\label{7}
\begin{array}{rl}
  H_{\rm SUSY} := & \displaystyle\frac{(-1)^{2s+1}}{2mc^2}\left( H^2-{\cal M}^2 \right)\\[4mm]
                = & \displaystyle\frac{1}{2mc^2} \left(\begin{array}{cc} AA^\dag& 0 \\ 0 & A^\dag A \end{array}\right)\geq 0
\end{array}
\end{equation}
and the corresponding complex SUSY charge
\begin{equation}\label{8}
  Q:=\frac{1}{\sqrt{2mc^2}} \left(\begin{array}{cc} 0 & A \\ 0 & 0 \end{array}\right)\,,\quad
  Q^\dag =\frac{1}{\sqrt{2mc^2}} \left(\begin{array}{cc} 0 & 0 \\ A^\dag & 0 \end{array}\right)\,.
\end{equation}
It is now obvious that these operators together with the $Z_2$-grading (or Witten) operator $W:=\beta$ obey the SUSY algebra
\begin{equation}\label{9}
\begin{array}{l}
  H_{\rm SUSY} =\{Q,Q^\dag\}\,,\quad \{Q,W\}=0\,,\quad Q^2=0=(Q^\dag)^2 \,,\\[2mm]
  \left[ W,H_{\rm SUSY} \right] = 0\,,\quad W^2 =1\,.
\end{array}
\end{equation}
Hence, an arbitrary spin Hamiltonian of the form (1) obeying the condition (5) may be called a {\it supersymmetric arbitrary-spin Hamiltonian}. This is consistent with the usual definition \cite{Thaller1992,Junker2019} of a supersymmetric Dirac Hamiltonian in the case $s=\frac{1}{2}$.
Let us note that condition (5) also implies that ${\cal M}$ commutes with all operators of the algebra (9).

It is interesting to note that the condition $[{\cal M}, {\cal O}]=0$ in addition implies that there exists an exact Foldy-Wouthuysen transformation \cite{Erik1958,Silenko,JunIno2018}
\begin{equation}\label{10}
  U:= \frac{|H|+\beta H}{\sqrt{2H^2+2{\cal M}|H|}}=\frac{1+\beta \,{\rm sgn} H}{\sqrt{2+\{{\rm sgn}H,\beta\}}}\,,
\end{equation}
$ {\rm sgn}H:=H/\sqrt{H^2}$, which brings the Hamiltonian into a block diagonal form, cf. eq.\ (6),
\begin{equation}\label{11}
  H_{\rm FW} := U H U^\dag = \beta \sqrt{H^2}=\beta |H|\,.
\end{equation}
As a side remark we mention that the two operators
\begin{equation}\label{10a}
\textstyle
  B_\pm :=\frac{1}{2}\left[1\pm \beta \right]\,,\quad \Lambda_\pm := \frac{1}{2}\left[1\pm {\rm sgn}H \right]
\end{equation}
are projection operators onto the subspaces ${\cal H}^\pm$ of positive and negative eigenvalues of $\beta$ and $H$, respectively, and they are related to each other via the unitary relation \cite{VZ1973}
\begin{equation}\label{10b}
  B_\pm = U \Lambda_\pm U^\dag\,.
\end{equation}
That is, the positive and negative energy eigenspaces are transformed via $U$ into eigenspaces of positive and negative eigenvalues of $\beta = W$, cf.\ eq.\ (\ref{11}). Note that we may express $U$ in terms of $B_\pm$ and $\Lambda_\pm $ as follows
\begin{equation}\label{UBL}
  U = \frac{B_+\Lambda_+ +B_-\Lambda_-}{\sqrt{(B_+\Lambda_+ +B_-\Lambda_-)(\Lambda_+B_+ +\Lambda_-B_-)}}\,.
\end{equation}

\section{Supersymmetric vector bosons}
Let us now consider the case of a vector boson with charge $e$ and mass $m$ interacting with a constant external magnetic field $\vec{B}$ characterised via the vector potential $\vec{A}=\frac{1}{2}\vec{B}\times\vec{r}$. The corresponding Hamiltonian is given by
\begin{equation}\label{12}
H=\left(\begin{array}{cc}
    M_+ & A \\ -A & -M_- \end{array}\right)\,,
\end{equation}
where
\begin{equation}\label{13}
  \begin{array}{l}
  \displaystyle
    M_\pm := mc^2 + \frac{\vec{\pi}^2}{2m}-\frac{ge\hbar}{2mc}(\vec{S}\cdot\vec{B})\,,\\[2mm]
  \displaystyle
    A := \frac{\vec{\pi}^2}{2m}-\frac{1}{m}(\vec{S}\cdot\vec{\pi})^2+\frac{(g-2)e\hbar}{2mc}(\vec{S}\cdot\vec{B}) =A^\dag
    \end{array}
\end{equation}
and $\vec{\pi}:= \vec{p}-e\vec{A}/c$.
Here $\vec{S}=(S_1,S_2,S_3)^T$ is a vector who's components are $3\times 3$ matrices obeying the $SO(3)$ algebra $[S_i,S_j]={\rm i} \varepsilon_{ijk}S_k$ representing the spin-one-degree of freedom of the particle, that is $\vec{S}^2=2$ as the spin $s=1$ for a vector boson. In above Hamiltonian we have introduced an arbitrary gyromagnetic factor $g$ which describes the coupling of this spin-degree of freedom to the magnetic field $\vec{B}$. Note that above Hamiltonian was, to the best of our knowledge, first derived in 1940 by Taketani and Sakata \cite{TS1940} with $g=1$. At the same time Corben and Schwinger \cite{CS1940} had studied the electromagnetic properties of mesotrons and concluded that $g=2$ is required to have a singularity free theory. For later work using above Hamiltonian with both $g=1$ and $g=2$ as well as arbitrary values  for $g$ see, for example, refs.\ \cite{YB1963,Guertin1974,Weaver1976,DF1992,DF1993,Silenko2014}.

Let us now investigate if above Hamiltonian (\ref{12}) together with (\ref{13}) does form a supersymmetric relativistic spin-1 Hamiltonian. For this we recall the relation \cite{DF1993}
\begin{equation}\label{14}
 \left[ \vec{\pi}^2,(\vec{S}\cdot\vec{\pi})^2\right] =\frac{2e\hbar}{c} \left[(\vec{S}\cdot\vec{B}),(\vec{S}\cdot\vec{\pi})^2 \right]\,
\end{equation}
which allows us to explicitly calculate the commutator
\begin{equation}\label{15}
  [M_\pm,A]=(g-2)\frac{e\hbar}{2m^2c}\left[(\vec{S}\cdot\vec{B}),(\vec{S}\cdot\vec{\pi})^2 \right]\,.
\end{equation}
Obviously for a non-vanishing magnetic field the SUSY condition (\ref{5}) is only fulfilled when $g=2$. In other words, when we require that the relativistic Hamiltonian for a massive charged spin-$1$ particle in a magnetic field is a {\it supersymmetric} Hamiltonian in above sense, only $g=2$ is allowed. This is indeed similar to the argument \cite{Junker2019} that the phenomenological non-relativistic Pauli-Hamiltonian for a charged spin-$\frac{1}{2}$ fermion exhibits a SUSY structure only when $g=2$.

\section{Concluding remarks}
In a final remark let us note that the above SUSY structure allows to reduce the eigenvalue problem of a supersymmetric arbitrary-spin Hamiltonian to that of a non-relativistic Hamiltonian $H_s$. As we will show elsewhere \cite{Junker2020}, for a charged particle in the constant magnetic field $\vec{B}$ the FW-transformed Hamiltonian (11) for the cases $s=0$, $s=\frac{1}{2}$ and $s=1$ takes the form
\begin{equation}\label{16}
  H_{\rm FW}=\beta mc^2\sqrt{1+\frac{2H_s}{mc^2}}\,,
\end{equation}
where
\begin{equation}\label{17}
  \begin{array}{rl}
    H_0    & := \displaystyle \frac{1}{2m}(\vec{p}-e\vec{A}/c)^2\,,\\[2mm]
    H_\frac{1}{2}& := \displaystyle \frac{1}{2m}(\vec{p}-e\vec{A}/c)^2-\frac{e\hbar}{mc}(\vec{\sigma}\cdot\vec{B})\,,\\[2mm]
    H_1    & :=\displaystyle \frac{1}{2m}(\vec{p}-e\vec{A}/c)^2-\frac{e\hbar}{mc}(\vec{S}\cdot\vec{B})\,.
  \end{array}
\end{equation}
Obviously, $H_0$ and $H_\frac{1}{2}$ represent the well-known non-relativistic Landau and Pauli-Hamiltonian, respectively, and $H_1$ is the correct non-relativistic Hamiltonian for a spin-1 particle in a magnetic field with $g=2$. In above $\vec{\sigma}$ is a vector whose components consist of Pauli's $2\times 2$ matrices representing the spin-$\frac{1}{2}$-degree of freedom.

The purpose of this short note is two-fold. First we generalised the concept of supersymmetric relativistic Hamiltonians known from the Dirac Hamiltonian with $s=\frac{1}{2}$ to the general case of arbitrary $s$. More explicitly, under the condition (\ref{5}) it was shown that a SUSY structure, cf.\ eqs.\ (\ref{7})-(\ref{9}), can be accommodated. Second, by considering a massive charged vector boson, i.e.\ $s=1$, in the presents of a constant magnetic field, this system resembles a SUSY structure if and only if its gyromagnetic factor $g=2$.

\acknowledgments
I have enjoyed enlightening discussions with Mikhail Plyushchay for which I am very grateful.

\end{document}